\documentclass[twocolumn,times]{aastex63}

\newcommand{\tar}{LP\,40$-$365}
\newcommand{\kms}{km\,s$^{-1}$}

\shorttitle{8.9-hr Rotation in LP\,40$-$365}
\shortauthors{Hermes et al.}

\begin{document}

\title{8.9-hr Rotation in the Partly Burnt Runaway Stellar Remnant LP\,40$-$365 (GD 492)}

\correspondingauthor{J. J. Hermes}
\email{jjhermes@bu.edu}

\author[0000-0001-5941-2286]{J.~J.~Hermes}
\affil{Department of Astronomy \& Institute for Astrophysical Research, Boston University, 725 Commonwealth Ave., Boston, MA 02215, USA}

\author[0000-0002-8935-0431]{Odelia~Putterman}
\affil{Department of Astronomy \& Institute for Astrophysical Research, Boston University, 725 Commonwealth Ave., Boston, MA 02215, USA}

\author[0000-0003-0089-2080]{Mark~A.~Hollands}
\affil{Department of Physics, University of Warwick, Coventry CV4 7AL,UK}

\author[0000-0001-9667-9449]{David~J.~Wilson}
\affil{McDonald Observatory, University of Texas at Austin, Austin, TX 78712, USA}

\author[0000-0001-6515-9854]{Andrew~Swan}
\affil{Department of Physics \& Astronomy, University College London, Gower Street, London WC1E 6BT, UK}

\author[0000-0002-9090-9191]{Roberto~Raddi}
\affil{Departament de F\'{\i}sica, Universitat Polit\`{e}cnica de Catalunya, c/Esteve Terrades 5, 08860 Castelldefels, Spain}

\author[0000-0002-9632-6106]{Ken~J.~Shen}
\affil{Department of Astronomy and Theoretical Astrophysics Center, University of California, Berkeley, CA 94720, USA}

\author[0000-0002-2761-3005]{Boris~T.~G{\"a}nsicke}
\affil{Department of Physics, University of Warwick, Coventry CV4 7AL,UK}

\begin{abstract}

We report the detection of $8.914$-hr variability in both optical and ultraviolet light curves of LP\,40$-$365 (also known as GD\,492), the prototype for a class of partly burnt runaway stars that have been ejected from a binary due to a thermonuclear supernova event. We first detected this $1.0$\% amplitude variation in optical photometry collected by the {\em Transiting Exoplanet Survey Satellite}. Re-analysis of observations from the {\em Hubble Space Telescope} at the {\em TESS} period and ephemeris reveal a $5.8$\% variation in the ultraviolet of this $9800$\,K stellar remnant. We propose that this $8.914$-hr photometric variation reveals the current surface rotation rate of LP\,40$-$365, and is caused by some kind of surface inhomogeneity rotating in and out of view, though a lack of observed Zeeman splitting puts an upper limit on the magnetic field of $<$$20$\,kG. We explore ways in which the present rotation period can constrain progenitor scenarios if angular momentum was mostly conserved, which suggests that the survivor LP\,40$-$365 was not the donor star but was most likely the bound remnant of a mostly disrupted white dwarf that underwent advanced burning from an underluminous (Type Iax) supernova.

\end{abstract}

\keywords{stars: individual: LP 40-365 --- stars: white dwarfs --- stars: rotation --- stars: kinematics and dynamics}

\section{Introduction}
\label{sec:intro}

\tar\ (a.k.a. GD\,492) is an exceptionally peculiar star: it is one of the most metal-rich stars known, with an atmosphere dominated by oxygen and neon that is also abundant in heavy elements from partial oxygen and silicon burning, with no detectable hydrogen or helium \citep{2017Sci...357..680V}. Most peculiarly, it is rapidly departing the Galaxy, with a rest-frame velocity of $v_{\mathrm{rf}}$\,$\simeq$\,$852$\,\kms\ that makes it unbound to the Milky Way; unlike hypervelocity stars \citep{2015ARA&A..53...15B}, \tar\ almost certainly did not come from the Galactic center \citep{2018MNRAS.479L..96R}. Currently, it appears to be an isolated star without radial-velocity variability, to a limit of roughly $2-3$\,\kms\ \citep{2018ApJ...858....3R}.

\tar, the prototype to a small but growing class of similar objects \citep{2019MNRAS.489.1489R}, derives these peculiarities because it was likely involved in a thermonuclear supernova event that ejected it from a short-period binary system. The original discoverers suggested that stars like \tar\ are most likely the bound remnants of partially disruptive thermonuclear supernovae, underluminous explosions often referred to as Type Iax supernovae \citep{2017Sci...357..680V}. Type Iax supernovae are thought to arise from an asymmetrically ignited deflagration in a Chandrasekhar-mass white dwarf that does not fully disrupt the star and leaves behind a bound remnant \citep[e.g.,][]{2012ApJ...761L..23J,2013ApJ...767...57F,2014MNRAS.438.1762F,2017hsn..book..375J}.

{\em Gaia} has enabled discovery of another class of runaways, the so-called ``dynamically driven double-degenerate double-detonation'' (D$^6$) stars, which are moving with such high velocities ($>$$1000$\,\kms) they must be the surviving donor stars in double white dwarf systems that underwent thermonuclear supernovae \citep{2018ApJ...865...15S}. In contrast, it is hard to constrain with certainty whether the \tar-like stars were the donor or accretor in a close binary, since detailed abundance analyses is complicated by highly uncertain metal diffusion times and nucleosynthetic yields. Most aspects of abundance analysis suggests that these objects are the bound remnants of the actual white dwarf in which the partial deflagration occurred \citep{2018ApJ...858....3R}. However, the \tar-like stars have also been analyzed as the polluted, inflated remnants of hot subdwarf donors in short-period progenitor systems \citep{2019ApJ...887...68B}.

One useful boundary condition that could help us constrain the progenitor system for \tar\ would be the final rotation rate of the stellar remnant. The highest-resolution spectroscopic observations of \tar\ have only put an upper limit on the rotation velocity ($v_{\rm rot} \sin{i} < 50$~km\,s$^{-1}$) that implies a rotation period longer than 4\,hr, based on current radius estimates \citep{2018ApJ...858....3R}.

We report here the significant detection of photometric variability in \tar\ at a period of $8.914$\,hr, present in an optical light curve collected by the {\em Transiting Exoplanet Survey Satellite} ({\em TESS}), with additional confirmation from an ultraviolet light curve from archival observations collected by the {\em Hubble Space Telescope} ({\em HST}) and possible detection in a near-infrared light curve from data collected by the {\em Wide-field Infrared Survey Explorer} ({\em WISE}).

We interpret this $8.914$-hr signal to correspond to the current surface rotation rate of \tar. Section~\ref{sec:observations} and Section~\ref{sec:analysis} of this Letter detail our observations and analysis, respectively. We conclude with a discussion of how the final rotation period we observe now in \tar\ can inform our understanding of the progenitor conditions that led to the formation of this partly burnt runaway stellar remnant. 

\section{Observations and Reductions}
\label{sec:observations}

\subsection{TESS Observations}

\tar\ (TIC\,198510602, $T$=15.5\,mag) was observed by {\em TESS} for nearly four months, in Sector~14 (2019~Jul~18 to 2019~Aug~14), Sector~15 (2019~Aug~15 to 2019~Sep~10), Sector~20 (2019~Dec~24 to 2020~Jan~20), and Sector~21 (2020~Jan~21 to 2020~Feb~18), all at 30-min cadence in the full-frame images (it was not targeted by any shorter, 2-min-cadence observations). The light curves for all four sectors were extracted using the python package \texttt{eleanor} \citep{2019PASP..131i4502F}, and analyzed using the \texttt{lightkurve} package \citep{2018ascl.soft12013L}. {\em TESS} utilizes a broad, red optical filter centered at roughly $7865$\,\AA.

\begin{figure}
    \centering
    \includegraphics[width=0.99\columnwidth]{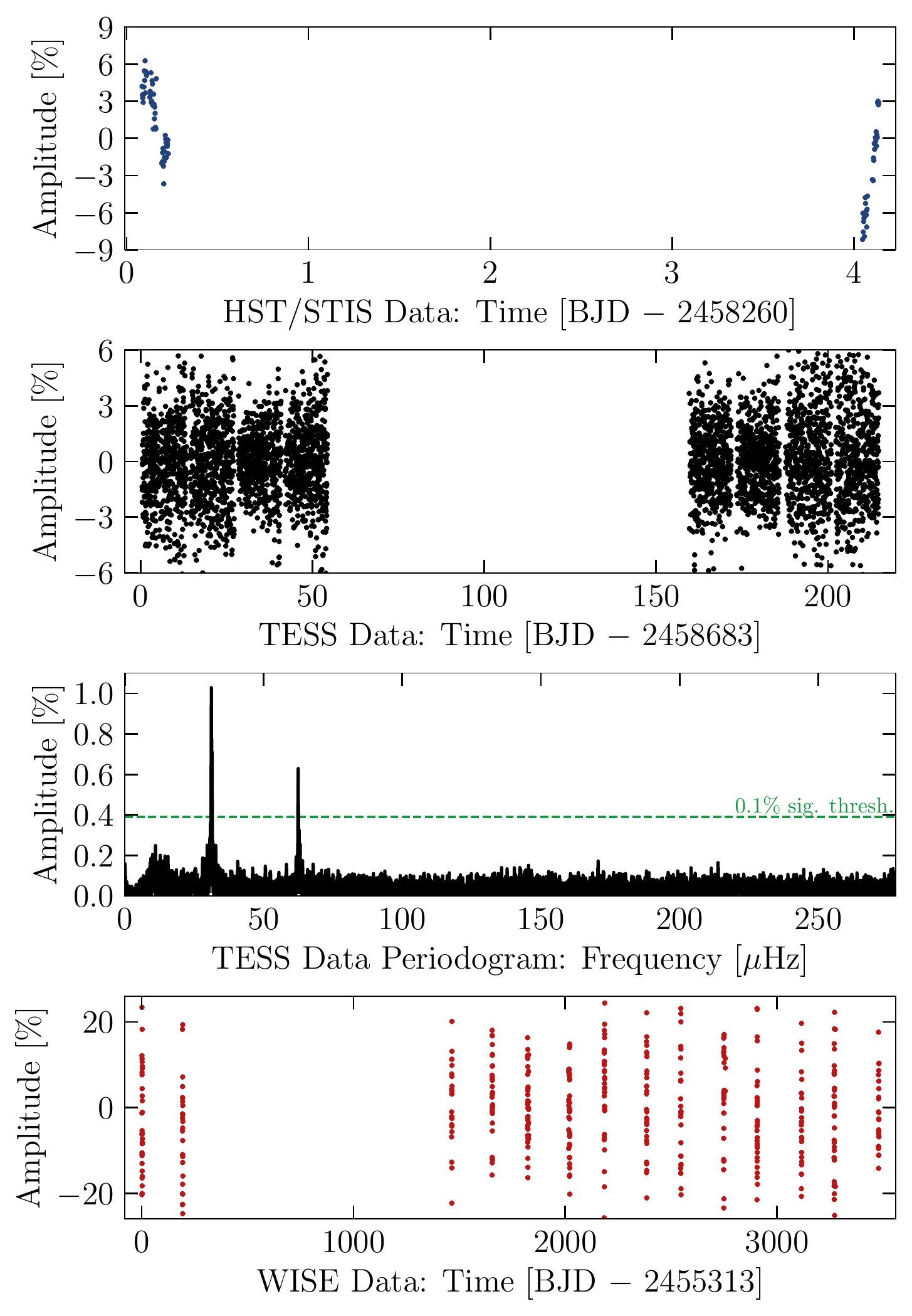}
    \caption{Full light curves of \tar\ used in this analysis. {\bf Top Panel:} Five orbits of {\em HST}/STIS ultraviolet observations from 2018~May. {\bf Second Panel:} Four sectors of {\em TESS} full-frame images from 2019~July to 2020~February. {\bf Third Panel:} Periodogram of {\em TESS} observations revealing a dominant period at $8.914$\,hr and its second harmonic. {\bf Bottom Panel:} More than 9.5 yr of {\em WISE} infrared photometry from 2010~April to 2019~November.}
    \label{fig:fullLC}
\end{figure}

The second panel of Figure~\ref{fig:fullLC} shows the full {\em TESS} light curve. Gaps in the data are present roughly every $13.7$\,days during each regular data download per orbit, and increased scatter due to reflected light is often present before the data downloads when the spacecraft is closer to Earth.

Our light curves were extracted using a fixed aperture, tuned for each sector depending on the spacecraft roll angle. We used a 3-px aperture for Sector~14, a 5-px aperture for Sectors~15~and~20, and a 2-px aperture for Sector~21. This changes the amount of extra light from background sources in the large pixel size ($>$21 arcsec pixel$^{-1}$) for each sector, which makes comparing the relative amplitude from each sector more difficult. Therefore, the relative flux in our full light curve in Figure~\ref{fig:fullLC} was normalized to the median flux for each sector. No amplitude corrections have been made for possible crowding in the large sky area covered by these apertures, which may result in a suppressed amplitude in the {\em TESS} bandpass caused by flux dilution. Our full {\em TESS} light curve has 4338\,points over 214.4\,days, for a roughly 42\% duty cycle.

\subsection{HST Observations}

We targeted \tar\ with the Space Telescope Imagining Spectrograph (STIS) instrument aboard {\em HST} over five orbits in 2018~May, which were first described and analyzed in \citet{2019MNRAS.489.1489R}. The observations were proposed to analyze the ultraviolet spectrum and better constrain the effective temperature of \tar, as well as improve our elemental abundance analysis.

We have reanalyzed these $15{,}253$ seconds of STIS observations taken with the G230L grating by taking advantage of their time-tagged photon collection \citep{1998PASP..110.1183W}. We interacted with these data using the python package \texttt{stistools}\footnote{\href{https://stistools.readthedocs.io/}{https://stistools.readthedocs.io}} available from the Space Telescope Science Institute. We binned the data into 3-min intervals, then summed the spectra of each in the wavelength range with the most signal-to-noise (S/N), $1700-3100$\,\AA. This effectively created an ultraviolet light curve with 3-min exposures and an effective central wavelength of $2520$\,\AA. We corrected the light curve to barycentric time using \texttt{Astropy} \citep{2013A&A...558A..33A}. The full {\em HST}/STIS light curve has 82\,points over 4.05\,days, for a roughly 4\% duty cycle, and is shown in the top panel of Figure~\ref{fig:fullLC}.

{\em HST} is in constant orbit around the Earth, with natural changes in the focus of instruments like {\em STIS} due to thermal evolution (often described as breathing, e.g., \citealt{2013MNRAS.436.2956S}). Since our {\em TESS} observations revealed variability at much longer timescales than the {\em HST} orbit we have not corrected for this effect. However, it likely has added increased scatter on timescales shorter than $95.4$\,min.

\subsection{WISE Observations}

We constructed an infrared light curve for \tar\ using data from the {\em Wide-field Infrared Survey Explorer} ({\em WISE}) spacecraft, which is performing an all-sky survey, visiting most sources twice per year \citep{2010AJ....140.1868W}. We queried the AllWISE Multiepoch Photometry Table and NEOWISE-R Single Exposure Source Table\footnote{Available at {\href{https://irsa.ipac.caltech.edu}{irsa.ipac.caltech.edu}}}, obtaining photometry in the \textit{W1} and \textit{W2} bands, taken during the decade 2010--2019. Here we only analyzed the higher-S/N observations in the \textit{W1} band, which has an effective wavelength of roughly $3.4$\,\micron.

The {\em WISE} scanning pattern resulted in on average 32 separate 7.7-second exposures per visit for \tar. Visits are spaced about six months apart, and last between two and eight days. Photometry is often analyzed after combining exposures within each visit (e.g., \citealt{2019MNRAS.484L.109S,2020ApJ...900...56S}), but we kept our observations unbinned, resulting in the relatively large scatter seen in the bottom panel of Figure~\ref{fig:fullLC}. Still, each exposure used has a formal S/N $>$$4$, with an average S/N of $9$. We corrected the light curve to barycentric time using \texttt{Astropy}.

The point spread function in the \textit{W1} band has a full-width-half-maximum of 6.1\,arcsec, and thus source confusion is a potential concern \citep{2020ApJ...891...97D}. However, no excess infrared flux is observed \citep{2018ApJ...858....3R}, and the median discrepancy between measured positions and those predicted from \textit{Gaia} astrometry is 0.3\,arcsec. The full {\em WISE}/\textit{W1} light curve has 451 points over 3480 days, for a roughly 0.001\% duty cycle, and is shown in the bottom panel of Figure~\ref{fig:fullLC}.

\section{Light Curve Analysis}
\label{sec:analysis}

A Lomb-Scargle periodogram of all four sectors of {\em TESS} data, shown in the third panel of Figure~\ref{fig:fullLC}, reveals two significant peaks: one dominant at roughly $8.90$\,hr and another peak at exactly half that period, roughly $4.45$\,hr. The 0.1\% significance threshold shown, at 0.39\% amplitude, was bootstrapped by keeping the time sampling identical but randomly shuffling the flux values in the light curve, noting the maximum peak for 99.9\% of all light curves from noise alone, as described in \citet{2015MNRAS.451.1701H}.

A simultaneous nonlinear least-squares fit of all {\em TESS} data, forcing the second frequency to be exactly twice that of the dominant signal, finds a best-fit period of $8.91405\pm0.00035$\,hr. We have performed the same weighted least-squares fit to the combined {\em TESS+WISE} dataset, which yields a more precise period, which we adopt throughout: $8.914122\pm0.000020$\,hr. From the first observed time of minimum in {\em TESS} we define the ephemeris as:

$$ \rm{BJD}_{\rm TDB} = 2458683.7076(23) + 0.37142177(83) \; E $$

We fix this period and compute a linear least-squares fit for each of the three light curves to define the amplitude of variability in each dataset. We find an ultraviolet amplitude of $5.79\pm0.21$\% from the {\em HST} data, an optical amplitude of $1.038\pm0.042$\% from the {\em TESS} data, and an infrared amplitude of $1.94\pm0.81$\% from the {\em WISE} data. The uncertainties on the amplitudes are purely formal, and do not include any systematic effects.

\begin{figure}
    \centering
    \includegraphics[width=0.99\columnwidth]{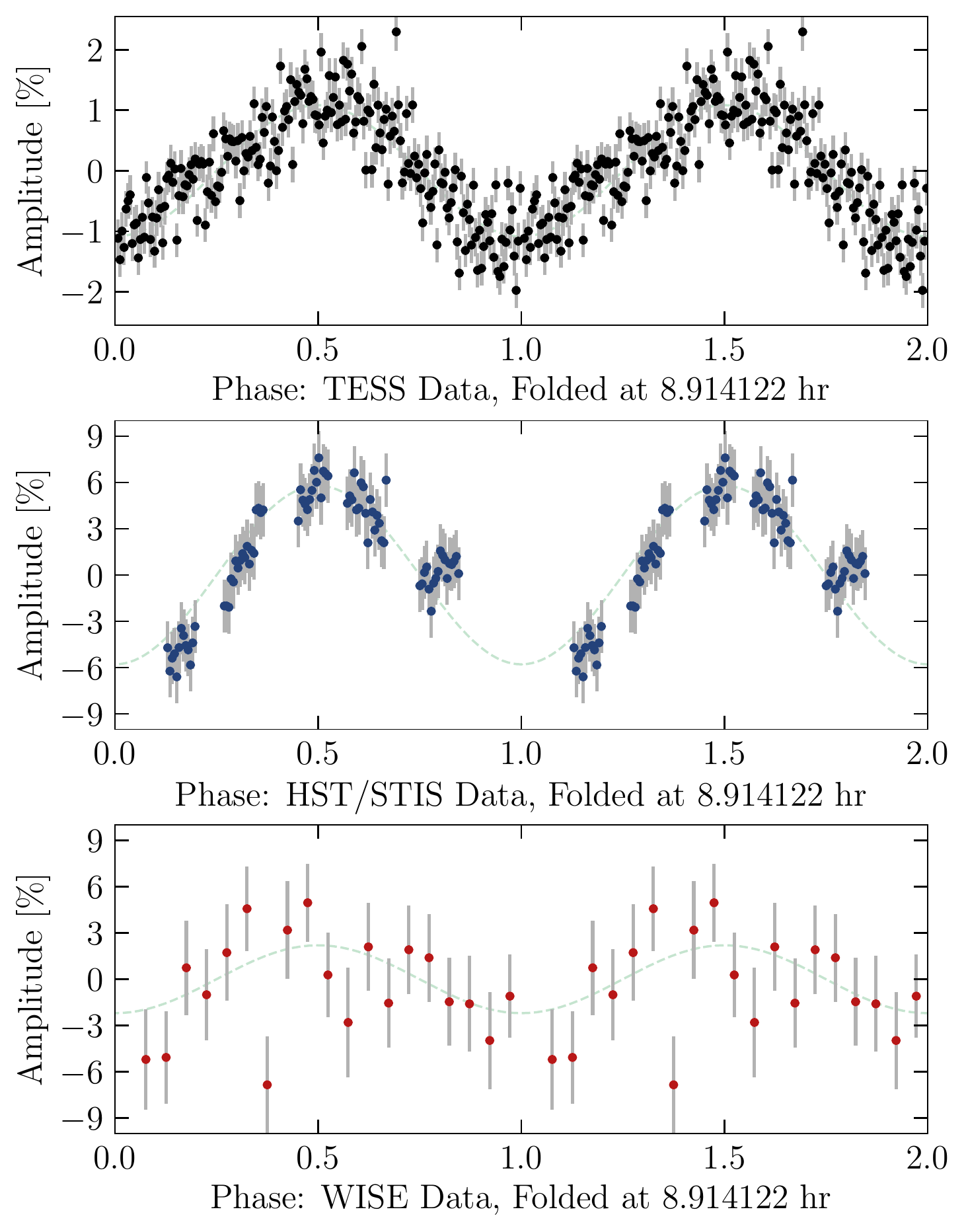}
    \caption{Folded light curves for \tar, with each phased using the ephemeris defined from the {\em WISE+TESS} observations and repeated for clarity. {\bf Top Panel:} All {\em TESS} data, which have an effective wavelength of roughly $7900$\,\AA, binned into 200 phase bins. {\bf Middle Panel:} All {\em HST}/STIS data, unbinned, which have an effective wavelength of roughly $2520$\,\AA. {\bf Bottom Panel:} All {\em WISE}/\textit{W1} data binned into 20 phase bins, which have an effective wavelength of roughly $3.4$\,\micron.}
    \label{fig:fig2}
\end{figure}

We use the ephemeris to phase all light curves to the same baseline, which we show visually in Figure~\ref{fig:fig2}. The best-fit amplitudes reported above are shown visually as a faint green dashed line in each panel. The phase-folded, binned {\em WISE} data is best fit with an amplitude larger than the full unbinned light curve ($2.20\pm0.87$\%).

The significant detection of the second harmonic of the $8.914$-hr fundamental period in the periodogram in Figure~\ref{fig:fullLC} suggests an asymmetry in the light curve. Non-sinusoidal light curves have been observed in other white dwarfs exhibiting photometric variability from surface spots (e.g., \citealt{2017MNRAS.468.1946H, 2017ApJ...835..277H, 2018AJ....156..119H}), and the second harmonic may be modeled in the future to constrain the inclination to the star. The folded light curves in Figure~\ref{fig:fig2} show that the photometric variations are mostly sinusoidal; excluding this harmonic does not change our reported ephemeris or measured amplitudes within the stated uncertainties.

We detect $8.914$-hr variability with significance in multiple datasets; the variability appears phase coherent over multiple years, as well as between the optical and ultraviolet wavelengths. In 2017 we observed \tar\ for 4\,hr from the 2-m Liverpool Telescope but saw no optical variability to a limit of roughly 0.3\% amplitude \citep{2018ApJ...858....3R}; it is likely the signal we see here was too long to detect from that light curve taken through a $V+R$-band filter. We cannot certify that the variability observed in the {\em WISE} observations is formally significant, although it is suggestive that the phased data in Figure~\ref{fig:fig2} shares the same photometric minimum as the {\em TESS} observations. Only 1\% of field sources vary at the wavelengths covered by \textit{W1}  \citep{2010ApJ...716..530K, 2018MNRAS.476.1111P}, so it is likely that the periodic variation we detect in {\em WISE} is from \tar.

We reserve a more detailed analysis of the amplitude ratios between the ultraviolet-optical-infrared datasets until we can properly assess crowding in the {\em TESS} observations, which could contribute to flux dilution and thus optical amplitude suppression. It is reasonable that the ultraviolet variations have higher amplitude than the optical variations, given that this is a 9800\,K stellar remnant  \citep{2019MNRAS.489.1489R}.

\section{Discussion and Conclusion}
\label{sec:conclusion}

We interpret the coherent $8.914$-hr photometric variability in \tar\ as the surface rotation period of this runaway stellar remnant. Spots caused by surface inhomogeneities on white dwarfs have been seen in a growing number of stars observed with space-based photometry from {\em Kepler} and {\em TESS} \citep{2017MNRAS.468.1946H}, especially in the presence of strong magnetic fields \citep{2013ApJ...773...47B}. Without knowing the topology of the magnetic field of this white dwarf, the inferred rotation period could be off by a factor of two; our inference of a $8.914$-hr rotation rate assumes a dipole field geometry.

We have reanalyzed the highest-resolution optical spectra collected from the R1200R grating from the 4.2-meter William Herschel Telescope on La Palma, described in \citet{2018ApJ...858....3R}. The sharp metal lines observed in that spectrum allowed us to put an upper limit on rotation of $v_{\rm rot}\,\sin{i}\,<\,50$~km\,s$^{-1}$ (we would infer a rotation velocity of roughly 22\,\kms\ from an $8.9$-hr rotation period). Similarly, using the observed shape of the calcium triplet between 8498--8662\,\AA\ (which is particularly sensitive to small magnetic fields), we do not detect any line-profile changes caused by Zeeman splitting from a magnetic field. This allows us to put an upper limit on the magnetic field in \tar\ of $<$\,$20$\,kG. Such a small field may not preclude formation of spots from surface inhomogeneities --- for example, we observe a spot from rotation in a pulsating white dwarf which has a global magnetic field $<$\,$10$\,kG \citep{2017ApJ...835..277H}.

The best estimates of the parameters of \tar\ are detailed in \citet{2019MNRAS.489.1489R}: $T_{\rm eff}=9800\pm300$\,K, $\log{g} = 5.5\pm0.3$\,[cgs], $L=0.20\pm0.04$\,L$_{\sun}$, $R=0.16\pm0.01$\,R$_{\sun}$, and $M=0.28^{+0.28}_{-0.14}$\,M$_{\sun}$. Using these atmospheric parameters, we find the gas pressure at an optical depth of $\tau=2/3$ is $10^{4.7}$\,dyne\,cm$^{-2}$. The magnetic pressure will exceed this gas pressure at roughly 1\,kG, so our 20\,kG upper limit allows for a field that can measurably influence the atmospheric structure at the photosphere.

It is likely that the partly burnt remnant \tar\ is still highly inflated; tracing back the kinematics yields an expected flight time of this star away from the Galactic disk of $5.3$\,Myr, which is considerably shorter than the Kelvin-Helmholtz timescale of $32$\,Myr. Considering this radius inflation, we have explored how the current rotation period can constrain progenitor scenarios if angular momentum in the system is mostly conserved.

We first consider if \tar\ was the white dwarf that underwent a thermonuclear event; in this first case we assume the progenitor star had a mass near the Chandrasekhar limit ($M_i$\,$>$\,$1.3$\,$M_{\sun}$), with a corresponding radius of $R_i<0.004$\,R$_{\sun}$  \citep{2005A&A...441..689A}. In this scenario, most of the mass from this star was lost in the thermonuclear event, and the radius has increased both because the lower-mass white dwarf is less degenerate but also because of eventual entropy release as the remnant evolves (e.g., \citealt{2019ApJ...872...29Z}). Assuming angular momentum transport is efficient and mostly conserved implies an initial rotation period of less than $2$\,min. This is very fast compared to typical isolated white dwarfs, which rotate with periods from $0.5-2$\,days \citep{2017ApJS..232...23H}, but is still much slower than breakup velocity (which would have a spin period of order seconds).

A $2$-min initial spin period is likely faster than the expected orbital period at explosion: \tar\ likely had an ejection velocity of roughly $v_{\rm ej} \simeq 600$\,\kms, which suggests a progenitor orbital period of roughly $10-30$\,min for most configurations \citep{2018MNRAS.479L..96R}. However, such spin-up would not be surprising for an accretor: spin-up is commonly expected (e.g., \citealt{1985A&A...148..207R}) and observed (e.g., \citealt{2020ApJ...898L..40L}) in cataclysmic variables with high mass-transfer rates onto accreting white dwarfs.

We also consider if \tar\ was the donor star ejected from a once-compact binary system; this scenario was modeled in detail by \citet{2019ApJ...887...68B}. They explored two different progenitors at different orbital periods: Their two hot subdwarf donor models (a 0.344\,$M_{\sun}$ donor onto a 0.927\,$M_{\sun}$ accretor and a 0.233\,$M_{\sun}$ donor onto a 0.779\,$M_{\sun}$ accretor) predict rapid rotation for the donor remnant $10$\,Myr after explosion; all donor remnants with final masses $>$$0.2$\,$M_{\sun}$ end up with rotation periods faster than roughly $1$\,hr, much faster than $8.9$\,hr. (Shocks and interactions with high ejecta kinetic energies will strip some or even most of the donor mass by the time it becomes a remnant, and the donor models specified above are pre-detonation masses.)

Combining the Keplerian orbital velocity of the donor with the expectations that it is Roche-lobe filling allowed \citet{2019ApJ...887...68B} to predict the radius of the donor at explosion in terms of its orbital velocity. The expected range of orbital velocities for \tar\ range from $600-800$\,\kms\ \citep{2019MNRAS.489.1489R}, which would predict a donor radius of $0.055-0.16$\,$R_{\sun}$ in orbital periods ranging from roughly $5-20$\,min.

The expected radius of the donor computed via the Roche-lobe radius depends on the unknown mass ratio of the system \citep{1983ApJ...268..368E} at explosion; ranges for $M_1$ are unlikely to exceed $0.2-0.5$\,$M_{\sun}$ for a core-He-burning subdwarf donor, and $M_2$ are unlikely to exceed $0.85-1.4$\,$M_{\sun}$ given that white dwarf detonations could be sub-Chandrasekhar in a subdwarf donor scenario \citep{2018ApJ...854...52S}. All told, we cannot explain the high ejection velocity with the comparatively slow rotation period currently observed in \tar\ if this is the donor remnant that was tidally locked in a short-period binary; the progenitor requires significantly more radius inflation to slow its rotation to the period we observe today.

There are two assumptions that could affect our inferences. It is likely a subdwarf donor would be close to tidally locked in these short-period binaries \citep{2018MNRAS.481..715P}, but breaking this assumption would affect initial spin-period estimates. Moreover, we cannot exclude a scenario in which the donor was highly inflated ($R$~$>$~$20$\,$R_{\sun}$) after explosion and lost significant angular momentum due to mass loss and magnetic braking. This phase should be relatively short-lived ($<$\,$10^5$\,years, e.g. \citealt{2013ApJ...773...49P}), but the mass loss may be significant enough to complicate the assumption of angular momentum conservation.

That said, if angular momentum is mostly conserved, we find it less likely that \tar\ represents the donor in a disrupted compact binary. Most likely, it is still experiencing a dramatic radius expansion best explained by it being the once-accreting white dwarf that itself underwent a thermonuclear event. This radius inflation is also supported by another much hotter \tar-like star, J1825$-$3757, which has a similar mass as \tar\ but a radius at least a factor of three larger \citep{2019MNRAS.489.1489R}. The event that disrupted the binary system caused both components to sling-shot apart at high speed. That thermonuclear event also led to partial burning in the progenitor star to \tar, as proposed by \citet{2017Sci...357..680V} and \citet{2018MNRAS.479L..96R,2018ApJ...858....3R}, based in part on the currently observed heavy-element abundances.

Clarifying the role of \tar\ in the explosion is important to further interpreting the abundances, especially to connect to expected nucleosynthetic yields (e.g., \citealt{2020ApJ...900...54L}). This connection is vital to help us better understand the peculiar properties of the class of partially burnt stellar remnants, the \tar\ stars \citep{2019MNRAS.489.1489R}.

\section*{Acknowledgements}
We acknowledge helpful comments from the anonymous referee, and thank Evan Bauer for valuable discussions. Support for this work was in part provided by NASA {\em TESS} Cycle 2 Grant 80NSSC20K0592. M.A.H. has received funding from the European Research Council under the European Union's Horizon 2020 research and innovation programme n. 677706 (WD3D). A.S. has received support from Science and Technology Facilities Council grant ST/R000476/1. R.R.\ has received funding from the postdoctoral fellowship program Beatriu de Pin\'os, funded by the Secretary of Universities and Research (Government of Catalonia) and by the Horizon 2020 program of research and innovation of the European Union under the Maria Sk\l{}odowska-Curie grant agreement No 801370.  K.J.S.\ was supported in part by NASA through the Astrophysics Theory Program (NNX17AG28G). B.T.G. was supported by the UK STFC grant ST/T000406/1 and by a Leverhulme Research Fellowship.

This paper includes data collected by the {\em TESS} mission. Funding for the {\em TESS} mission is provided by the NASA's Science Mission Directorate. Based on observations made with the NASA/ESA {\em Hubble Space Telescope}, obtained from the Data Archive at the Space Telescope Science Institute, which is operated by the Association of Universities for Research in Astronomy, Inc., under NASA contract NAS 5-26555. These observations are associated with program \#15431, and funding in-part provided by programs \#15871 and \#15918. This publication makes use of data products from the {\em Wide-field Infrared Survey Explorer}, which is a joint project of the University of California, Los Angeles, and the Jet Propulsion Laboratory/California Institute of Technology, funded by the National Aeronautics and Space Administration.


\end{document}